\documentclass[fleqn,12pt]{wlscirep}
\usepackage[utf8]{inputenc}
\usepackage[T1]{fontenc}
\usepackage{bm}
\usepackage{amsmath}
\usepackage{float}
\usepackage{booktabs}
\usepackage{multirow}
\title{\textit{Ab initio} Studies of Magnetism and Topology in solid Pd-rich $\bm{a}$-PdSi Alloys}

\author[1,+]{Isa\'{i}as Rodr\'{i}guez}
\author[2,+]{Renela M. Valladares}
\author[2,+]{Alexander Valladares}
\author[1,+]{David Hinojosa-Romero}
\author[1,+,*]{Ariel A. Valladares}
\affil[1]{Instituto de Investigaciones en Materiales, Universidad Nacional Autónoma de México, Apartado Postal 70-360, Ciudad Universitaria, CDMX, 04510, México.}
\affil[2]{Facultad de Ciencias, Universidad Nacional Autónoma de México, Apartado Postal 70-542, Ciudad Universitaria, CDMX, 04510, México.}

\affil[*]{valladar@unam.mx}

\affil[+]{these authors contributed equally to this work}


\begin{abstract}
In 1965 Duwez \textit{et al.} reported having generated an amorphous, stable phase of palladium-silicon in the region 15 to 23 atomic percent (at. \%) silicon. These pioneering efforts have led to the development of solid materials that are now known as Bulk Metallic Glasses (BMG). In 2019 we discovered, computationally, that bulk amorphous Pd becomes magnetic, and so does porous/amorphous Pd. Puzzled by our results we undertook the study of several solid binary systems in the Pd-rich zone; in particular, the study of the glassy metallic alloy $a$-Pd$_{100-c}$Si$_{c}$, for $0 \leq c \leq 22$, ($c$ in at. \%) to see what their topology is, what their electronic properties are and to inquire about their magnetism. Here we show that this metallic glass is in fact magnetic in the region $0 \leq c < 15$. Collaterally we present $\alpha$ and $\beta$ magnetization curves that manifest the net magnetic moment observed. We also discuss the topology and the position of the first few peaks of the pair distribution functions, which agrees well with experiment. The BMGs produced experimentally so far are limited in size, but despite this limitation, recent industrial efforts have developed some useful devices that may revolutionize technology. 
\end{abstract}
\begin{document}

\flushbottom
\maketitle

\thispagestyle{empty}

\section*{Introduction}

Ever since Klement, Jr. \textit{et al.} generated an amorphous, unstable, phase of a gold-silicon alloy in 1960 \cite{Klement_1960}, Au$_{75}$Si$_{25}$ in atomic percent (at. \%), much has been written and even more has been done in the field of glassy metals. In September 3, 1960, Pol Duwez and his two graduate students W. Klement, Jr., and R. H. Willens reported that by rapid solidification of the liquid, a non-crystalline structure of AuSi could be generated.  Thereafter, in 1965, Duwez \textit{et al.} \cite{Duwez_1965} obtained stable, amorphous metallic alloys of palladium and silicon, $a$-Pd$_{100-c}$Si$_{c}$ for concentrations $15 < c < 23$, using the same experimental approach as before; i.e., a rapid cooling from the melt. These are the beginnings of the production of glassy metals by rapid cooling from the liquid and the evolution of the field of Bulk Metallic Glasses (BMGs) was under way. The Pd-Si system is a prototypical, simple, example of this field.

In Duwez words, ``in September 1959, …  as part of a research program whose purpose was far remote from metallic glasses, an alloy containing 75 at. \% Au and 25 at. \% Si rapidly quenched from the liquid state appeared to be amorphous.'' \cite{Duwez_Guntherodt_book_1981}. A curiosity at first, with time it has become clear that glassy metals in general, and BMG in particular, have fascinating and potentially very useful properties \cite{Suryanarayana_Book_2018}. For example, some of the spectacular properties deal with their resistance to wear, that allows the use of them in lasting gears with several useful applications, like in the food industry where the use of lubrication may lead to the contamination of the products. Spin-offs of NASA are working to develop gears for space modules subject to extreme weather conditions that restrict the use of common lubricants \cite{Amorphology, Hofmann_2016}. Their mechanical properties are also worth mentioning since they are very resistant to stresses \cite{Suryanarayana_Book_2018} and therefore more durable. Imagination is the limit and so is the small size of the BMGs produced so far, since the largest specimen generated is a BMG of Pd$_{42.5}$Cu$_{30}$Ni$_{7.5}$P$_{20}$ with dimensions no larger than 10 cm along any of the three spatial directions \cite{Nishiyama_2012}.

Glass has been known for millennia \cite{Handbook_2019}, but it was during the last century that metallic glasses began to appear and to claim their place in the scientific and technological scenario. However, since the science of glass is even more recent, it should not surprise anyone the limited knowledge of some of the properties of these materials; it is now that we are beginning to understand why metallic glasses behave the way they do and the potential usefulness of amorphous solids in general. In the PdSi system, the oldest stable glassy binary prepared from the melt, Si is known as the glass forming element and ever since the amorphicity of this system was reported, studies have been conducted to understand their behavior. The PdSi alloys have been largely studied but there are features not well understood and some others to be researched.

The range of concentration considered by Duwez and coworkers, although very restricted, is illustrative enough to detonate the growth of diverse studies related to its properties and structure. The phase diagram for the palladium-silicon alloys indicates the presence of a eutectic structure at about 15 at. \% Si and at a temperature of the order of 1090 K, and the amorphous alloys obtained range in concentrations from 15 to 23 at. \% Si. It was Cohen and Turnbull that first pointed out that the proximity of the eutectic point was relevant to the formation of the amorphous structure \cite{Cohen_1961, Turnbull_1969} and from there on people started to look for eutecticity in phase diagrams to generate new amorphous glassy metallic alloys. The relevance of this result is that for the first time, an amorphous material could be formed by very rapid cooling from the melt, unlike other processes known at the time, and it was found that for the 20 at. \% Si alloy, undercooling as large as 300 °C could be reached \cite{Duwez_1965}. 

Despite obstacles, efforts continue to generate larger samples of BMGs to make them applicable in some everyday situations; however, scientific progress is slower than technology demands. When large samples of BMGs become available the technological possibilities will flourish, and this is the quest in many laboratories around the world.

\subsection*{Motives}

\begin{figure}[ht]
\centering
\includegraphics[width=\linewidth]{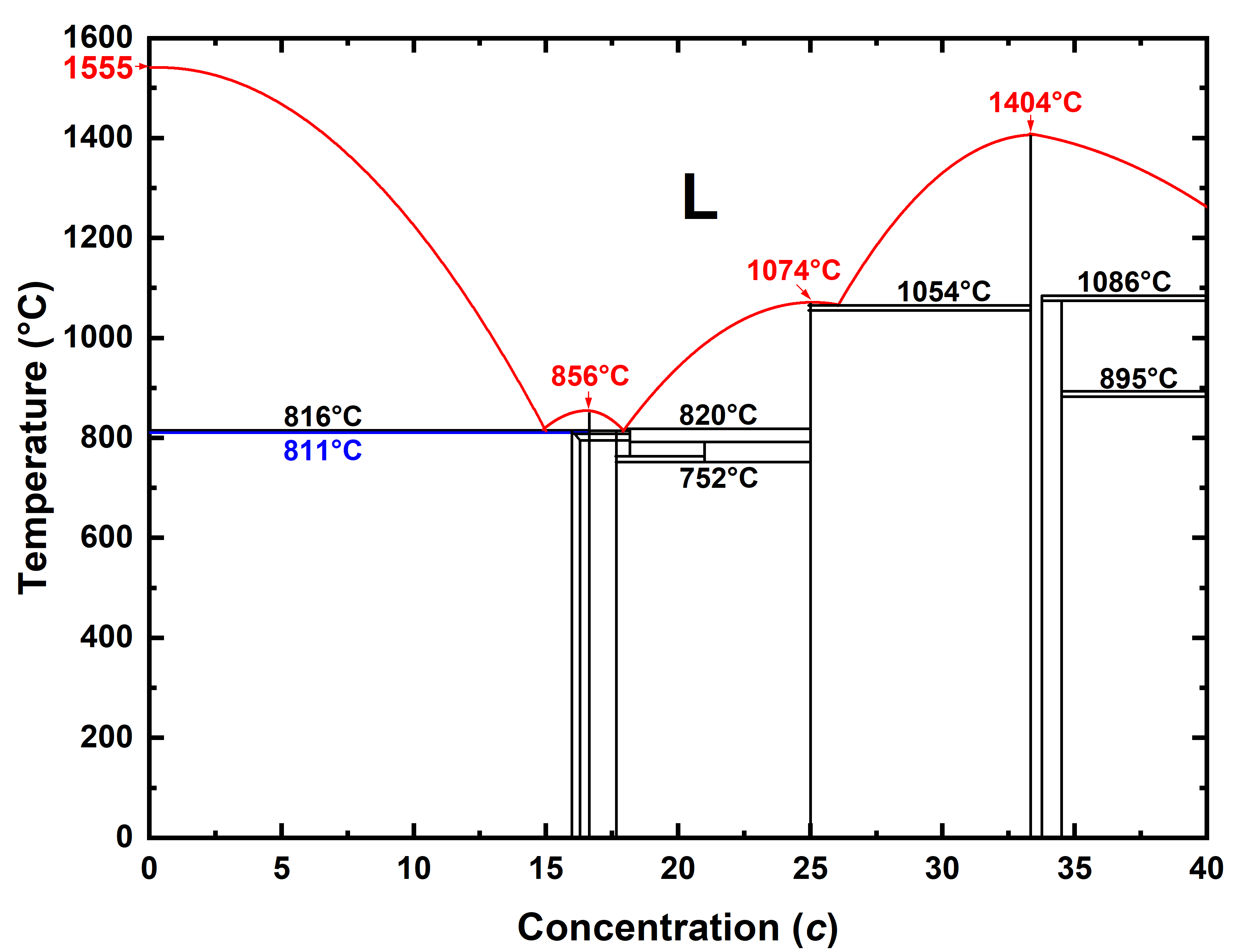}
\caption{Region of the Pd-Si phase diagram near the eutectic point. The atomic concentrations deployed are $0 \leq c \leq 40$ at. \%. This figure is a mathematical adjustment to the experimental values contained in References \cite{Okamoto_1993, Baxi_1991, Massara_1993}.}
\label{fig:Fig1}
\end{figure}

In 2019 we discovered that bulk amorphous Pd becomes magnetic \cite{Rodriguez_2019}. Puzzled by our results we undertook the task to study some Pd-based amorphous materials for concentrations close to the palladium-rich zone to see if this magnetism would persist, and to what extent. Contaminating $a$-Pd with hydrogen, deuterium or tritium to generate palladium ``ides'', $a$-Pd (H/D/T)$_{x}$, would be a natural path since Pd is well known for ad- and ab-sorbing hydrogen and its isotopes; to the point that it has been considered as an alternative to store H and use it in electrical vehicles. So we did the contamination and found that, in fact H, D, and T contribute to decrease the magnetism of amorphous Pd until, for values of the ratio $x$ close to 50 \% the magnetism completely disappears, and \textit{voilà} superconductivity appears giving rise to the so-called inverse isotope effect for the three isotopes, H, D and T \cite{Rodriguez_supercond_2021}. It was a fortunate circumstance that the Pd-rich zone for the $a$-PdSi is near a eutectic point that would explain, according to reference \cite{Cohen_1961}, the appearance of amorphous structures. In Figure \ref{fig:Fig1}, which is a mathematical adjustment of the experimental values found in References \cite{Okamoto_1993, Baxi_1991, Massara_1993}, we depict this region.

Magnetism in the liquid phase of palladium-silicon alloys has been studied by M\"{u}ller \textit{et al.} back in 1978 \cite{Muller_1978}. They found that for all silicon concentrations magnetism appears and the magnetic susceptibility decreases with increasing silicon and becomes negative for concentrations larger than 20 \%. At $c \approx 60 \%$ it becomes positive again and remains positive. They argue that due to the empirical similarity between glassy and liquid metals, magnetic inferences can be made concerning the solid glassy phases, and that therefore magnetism in the amorphous solid samples should appear in a similar manner and with a similar behavior as found for the liquid. However, no experimental, simulational or theoretical results have been found by the present authors, prior to our work. S\"{a}nger \cite{Sanger_1984} in 1984 analyzed the experimental results for the liquid and decomposed magnetism in para- and dia-, offering an explanation for the found, as will be shown later on.

But what about solid glassy metallic PdSi alloys? The results of magnetism for the liquid alloys, plus the results we have found for amorphous Pd, for porous/amorphous Pd, and for palladium hydrides, led us to the investigation of magnetism in these alloys. As far as we know, no work in the literature reports possible magnetic properties for the Pd-rich concentrations of the solid glassy palladium-silicon alloys. Also, we look at their electronic properties and report the densities of states for $\alpha$ and $\beta$ spins. Since there seems to exist a discrepancy in the experimentally reported position of the maxima of the first few peaks in the Pair Distribution Functions, PDFs, and some simulated results \cite{Fukunaga_1981, Wong_Guntherodt_book_1981}, we also look at these parameters. The study was conducted for $a$-Pd$_{100-c}$Si$_{c}$, with concentrations in the interval $0 \leq c \leq 22$. This is what we report in this paper.

\section*{Methods}

To generate the amorphous Pd-Si samples we used the \textit{undermelt-quench} procedure, a molecular-dynamics approach developed in our group that has led to very good specimens of the amorphous phases \cite{Rodriguez_2019, Valladares_2008}. Alloys with 8 different concentrations of silicon ($c$ = 2.5, 5, 10, 13.34, 15, 17.5, 20 and 22) were randomly arranged in a (non-stable) diamond-like supercell containing a total of 216 atoms of both elements. Care was exercised to construct these supercells using the experimental densities reported in the literature \cite{Okamoto_1993, Fukunaga_1981, Louzguine_Luzgin_2012}. Then using CASTEP \cite{Clark_2005} included in the Materials Studio suite of codes \cite{Biovia} we performed Molecular Dynamics processes, MDs, starting from the diamond-like non-equilibrium structures to propitiate the randomization of the structures. Once the MDs cycles were completed and the amorphization procedure finished, we geometry optimized, GO, the atomic structures looking for the topology that would locally minimize the energy. Clearly, the amorphous arrangement is not the \textit{minimum-minimorum} of the energy; such minimum energy structures would be the crystalline one. In this manner, when the final atomic distributions were attained, the structures were \textit{locally} stable and amorphous.

\subsection*{Specifics}

The code CASTEP of the Materials Studio suite of codes was utilized for all computational procedures, both MD and GO.

For the MD processes the following parameters were used: An NVT ensemble and the Nose-Hoover thermostat for the disordering thermal processes that consist of a heating ramp of 100 steps starting from 300 K and going up to 1500 K for $c < 10$ at \%, and a heating ramp of 100 steps starting from 300 K going up to 1000K for $c > 10$ at \%, staying always below the liquidus temperature. After the heating ramps, cooling ramps (with the same (absolute value) slope as the heating) of 125 steps were performed from the max temperature to 12 K (or 7K); in this manner disordering the structures was accomplished. The GGA-PBE functional was used in the process \cite{Perdew_1996}. A 300 eV cut-off energy for the plane-wave basis, a grid-scale of 2.0 for the energy minimization, and a Pulay mixing scheme were employed, with a thermal smearing of 0.1 eV for the occupation, together with a SCF energy threshold of $2.0 \times 10^{-6}$ eV.

For the GO of the amorphous structures, the following parameters were used:  We worked with the GGA-PBE functional also \cite{Perdew_1996}, with a plane-wave basis of 330 eV for the cut-off energy, a grid-scale of 2.0 and a fine-grid-scale of 3.0. For the energy minimization a Pulay mixing scheme was applied, with a 0.1 eV thermal smearing for the occupation together with an SCF energy threshold of $2.0 \times 10^{-6} eV$. The total spin of the specimens was not fixed during the heating and the cooling procedure, and neither during the GO processes, so the final structures were obtained with the spin unrestricted to allow for the evolution dictated by the interactions and the procedure. The time for each step was 7 fs and the total time for a typical heating and cooling cycle was 1.57 ps.

The SCF energy threshold was $5.0 \times 10^{-7}$ eV and for the BFGS minimizer (using delocalized internals) the following tolerances were employed: energy tolerance of $5.0 \times 10^{-7}$ eV, force tolerance of $1.0 \times 10^{-2}$ eV \AA$^{-1}$, and a maximum displacement of $5.0 \times 10^{-4}$ \AA.

\subsection*{Calculations}

At the end of the MD and GO processes the spin up ($\alpha$ spins) and spin down ($\beta$ spins) were determined to investigate the possible magnetism of these alloys. All results are reported in the next section. A collateral product of this investigation is the comparison of the positions of the first few prominent peaks of the PDFs with some experimental results obtained several decades ago \cite{Wong_Guntherodt_book_1981} and recently \cite{Ohkubo_2003}. At the time, there were discrepancies between the experimental values measured and those obtained in some simulations. 

\section*{Results}

\begin{figure}[H]
\centering
\includegraphics[width=0.7\linewidth]{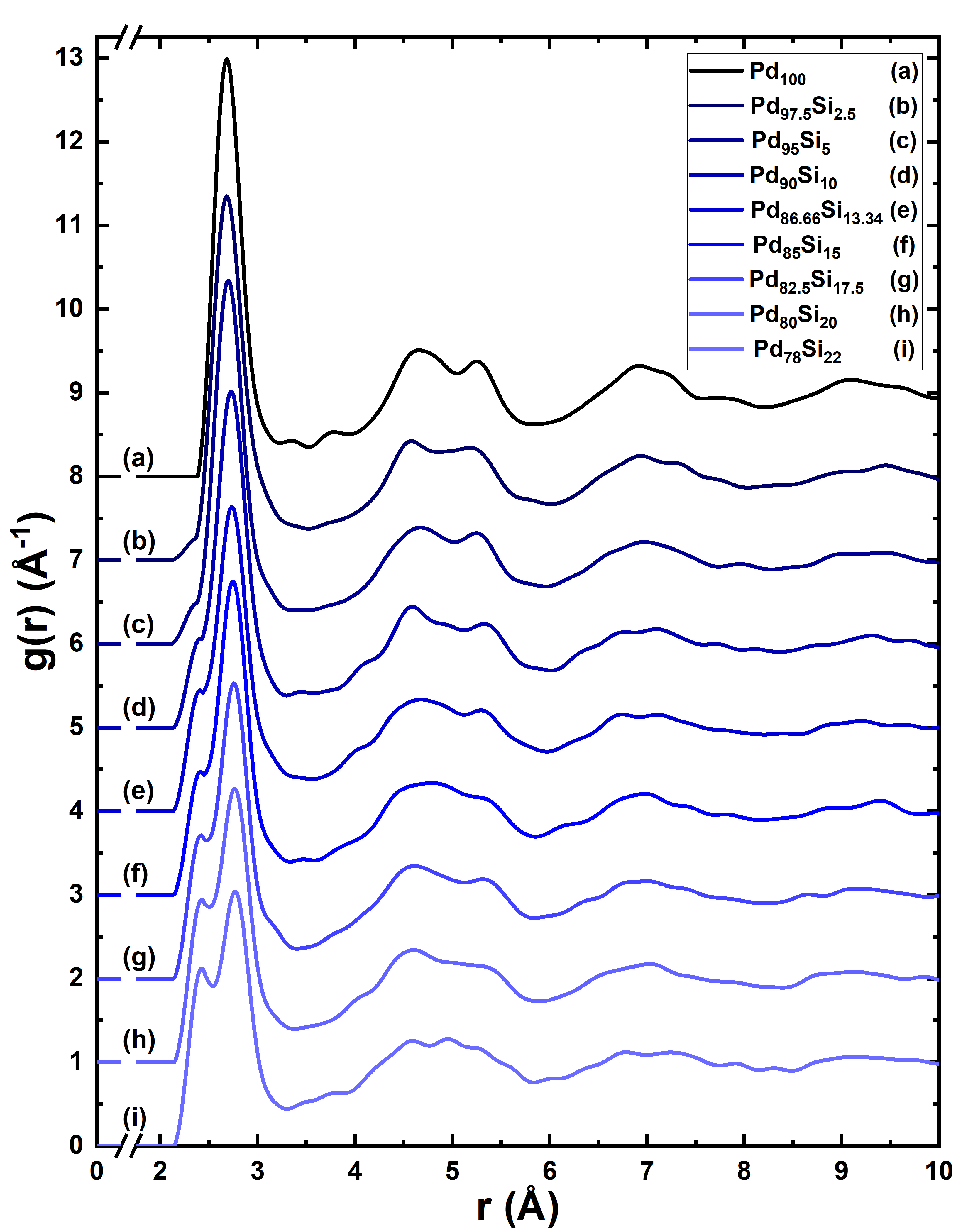}
\caption{Total Pair Distribution Functions, tPDFs, for the 8 alloys studied in this work and for the pure bulk palladium sample. The bimodal character of the second peak, typical of amorphous Pd, gradually disappears as the silicon concentration increases. The tPDFs were calculated using \texttt{Correlation}, an open-source software developed by Rodr\'{i}guez \textit{et al.} \cite{Correlation_2021}.}
\label{fig:Fig2}
\end{figure}

\begin{figure}[H]
\centering
\includegraphics[width=0.85\linewidth]{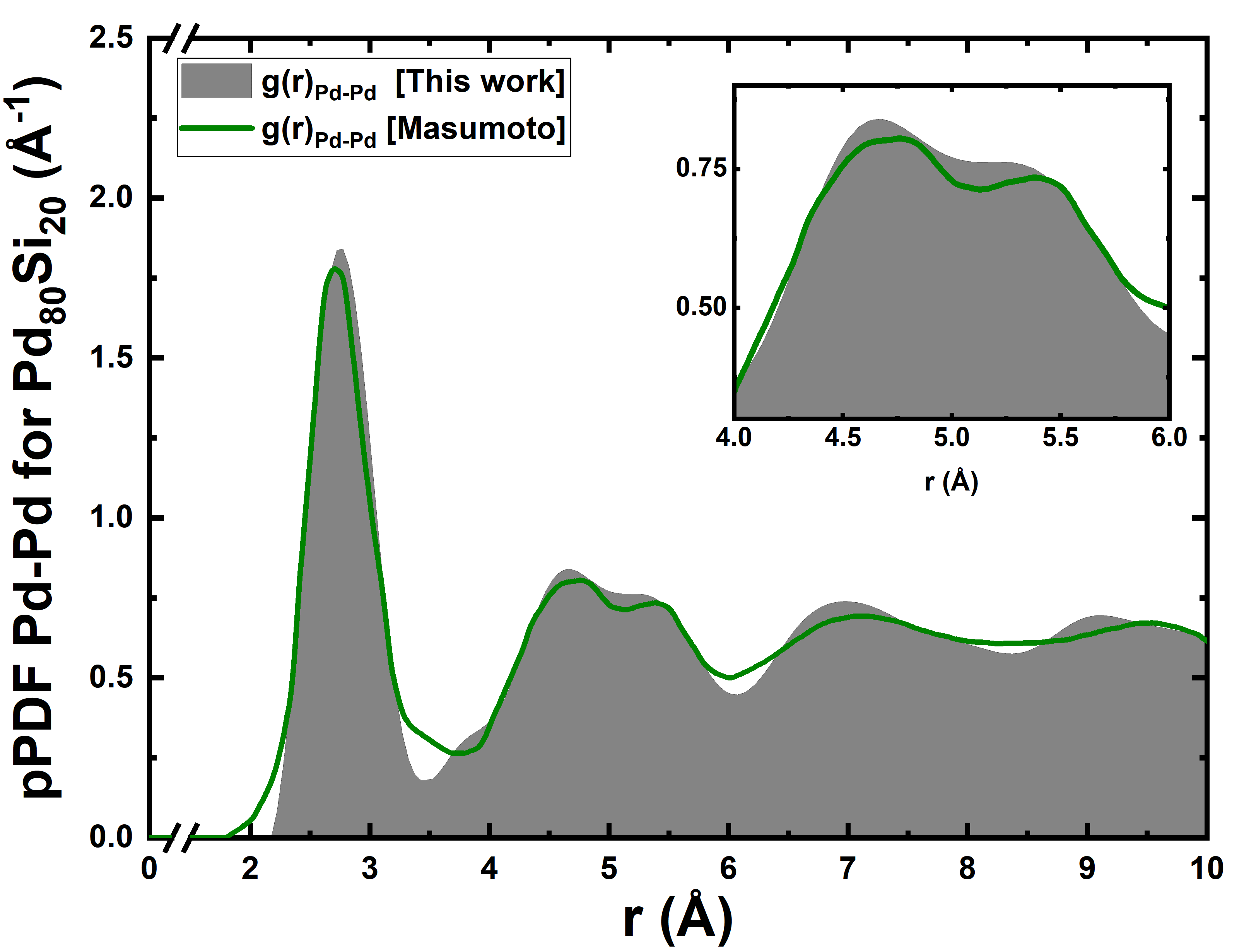}
\caption{Comparison of the Pd-Pd pPDF for the Pd$_{80}$Si$_{20}$ sample with the partial experimental result of Masumoto (in Waseda's book \cite{Waseda_1980}) \cite{Waseda_1975}. In the inset notice the bimodal nature of the second peak, reminiscent of the bimodal shape of the snake profile that swallowed an elephant in Le Petit Prince \cite{The_Little_Prince}.}
\label{fig:Fig3}
\end{figure}

Figure \ref{fig:Fig2} represents the total Pair Distribution Functions, tPDFs, of the nine supercells studied in this work, 8 amorphous alloys plus the pure amorphous palladium sample \cite{Rodriguez_2019}. The 216-atom initially unstable supercells have a diamond-like structure with densities determined by experiment \cite{Okamoto_1993, Fukunaga_1981, Louzguine_Luzgin_2012} and an edge length that goes from 14.5749 \AA\ for the Pd$_{78}$Si$_{22}$ to 14.7058 \AA\ for the Pd$_{100}$ sample. PDFs are difficult to obtain experimentally but are the best global description of the amorphous atomic topology of pure elements and alloys. In particular, the partial PDFs, pPDFs, require more labor by the experimentalists and they are not as frequently reported as the total. In our approach we can obtain partial and total PDFs. In a previous paper of our group \cite{Alvarez_2003} we contrasted the assumptions that experimentalists have to invoke to describe partial PDFs and showed that our approach is much better than to assume Gaussian curves fitted to the position of the peaks observed in the PDFs. The agreement of the silicon low concentration tPDFs with the experimental results included in Ref. \cite{Rodriguez_2019} for pure amorphous Pd is to be noted, and indicates consistent results in our simulations, see Figure \ref{fig:Fig3}.

\begin{figure}[H]
\centering
\includegraphics[width=0.85\linewidth]{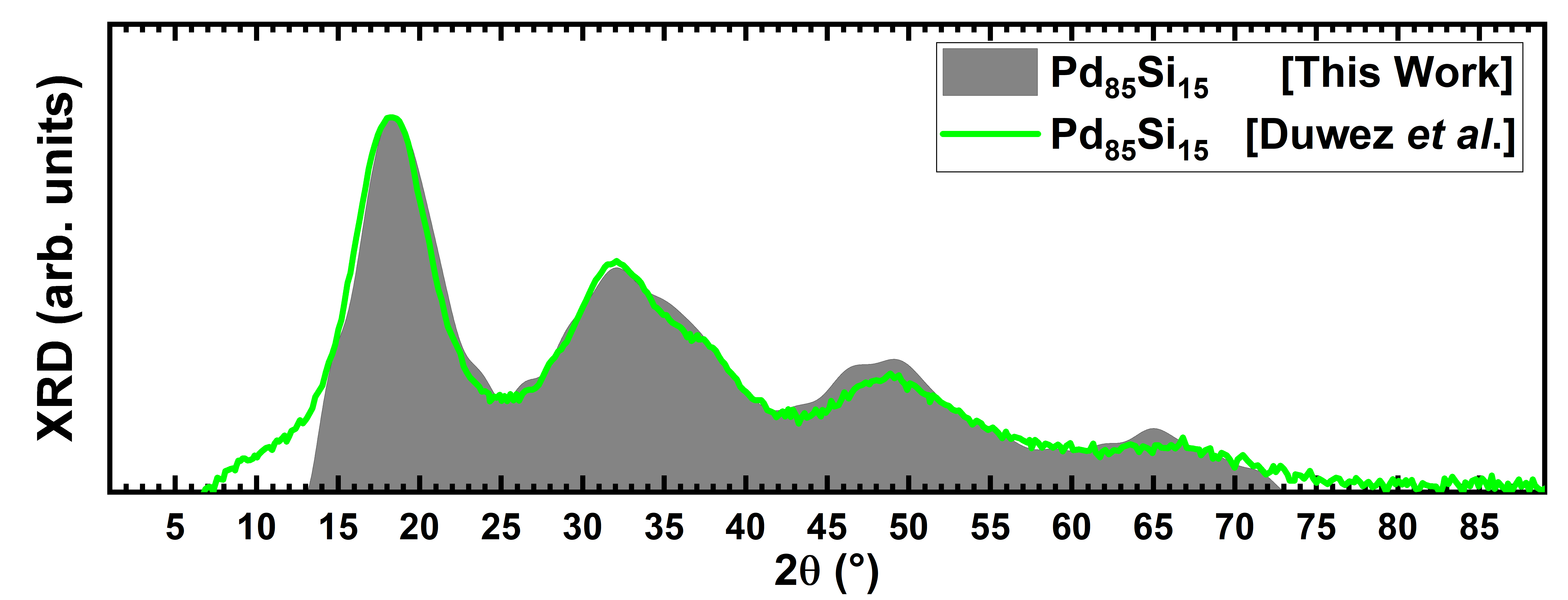}
\caption{Comparison of the XRD obtained from the experimental results of Ref. \cite{Duwez_1965} ($c = 15$ at. \%), green line, with our simulations, dark solid profile. See text.}
\label{fig:Fig4}
\end{figure}

To compare with the results of Duwez \textit{et al.} \cite{Duwez_1965}, we calculated the XRD using Reflex, a package included in the Materials Studio suite of codes. Considering the fact that the units are arbitrary we superimposed the first peak of our simulation with the first peak of the experimental results and then both curves reasonably coincided for most of the angles considered in the experiment, Figure \ref{fig:Fig4}. Since Duwez and co-workers did this XRD study for $a$-PdSi with a silicon concentration of 15 \% the comparison is done with one of our samples constructed for the same concentration.

\begin{figure}[H]
\centering
\includegraphics[width=0.7\linewidth]{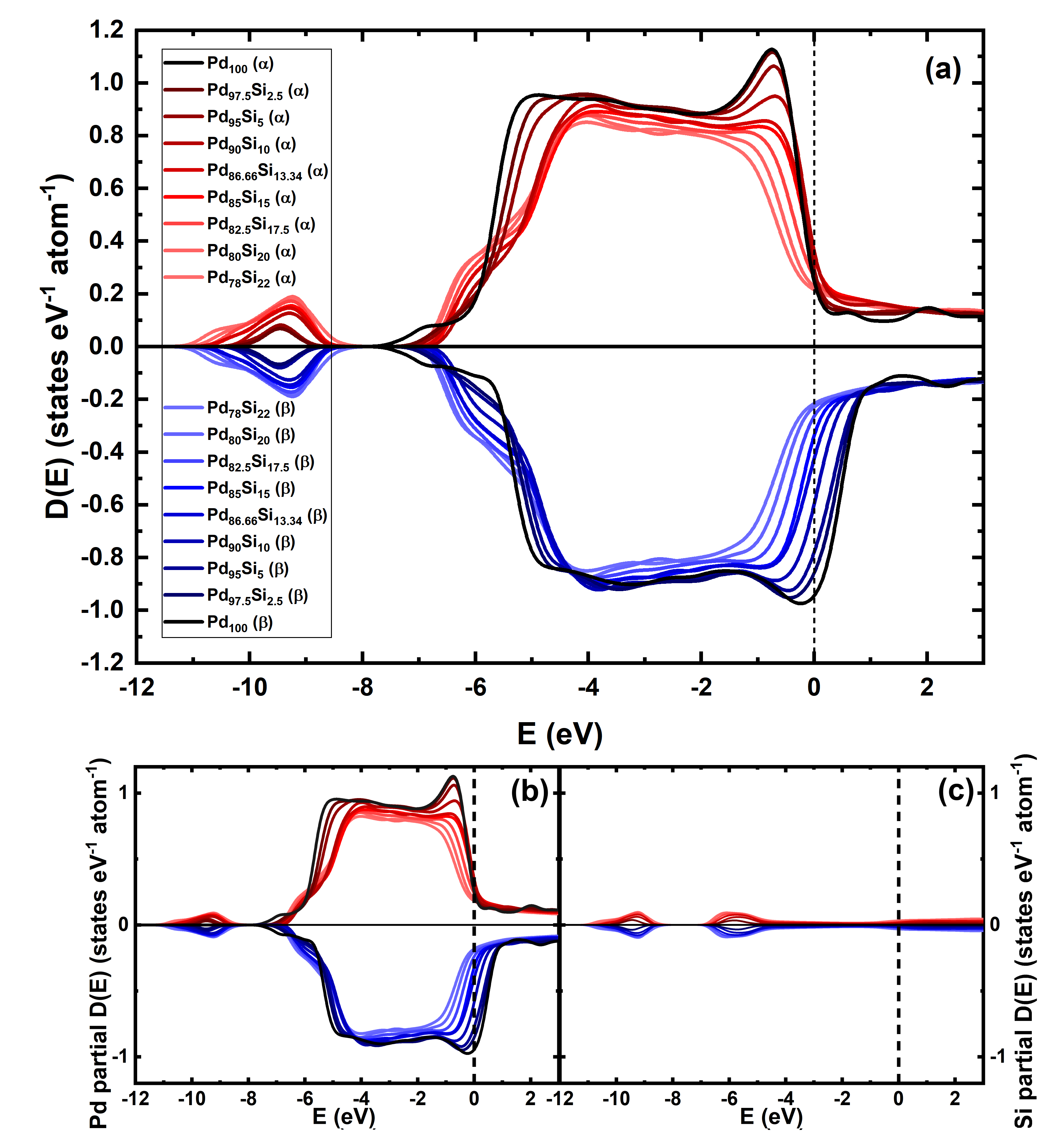}
\caption{Spin up ($\alpha$ spins) and spin down ($\beta$ spins) densities of states for the 9 supercells. (a) Total densities of states (b) Palladium partial densities of states (c) Silicon partial densities of states. The \textbf{\textit{net}} magnetic moment tends to zero as the silicon concentration increases.}
\label{fig:Fig5}
\end{figure}

\begin{figure}[H]
\centering
\includegraphics[width=0.85\linewidth]{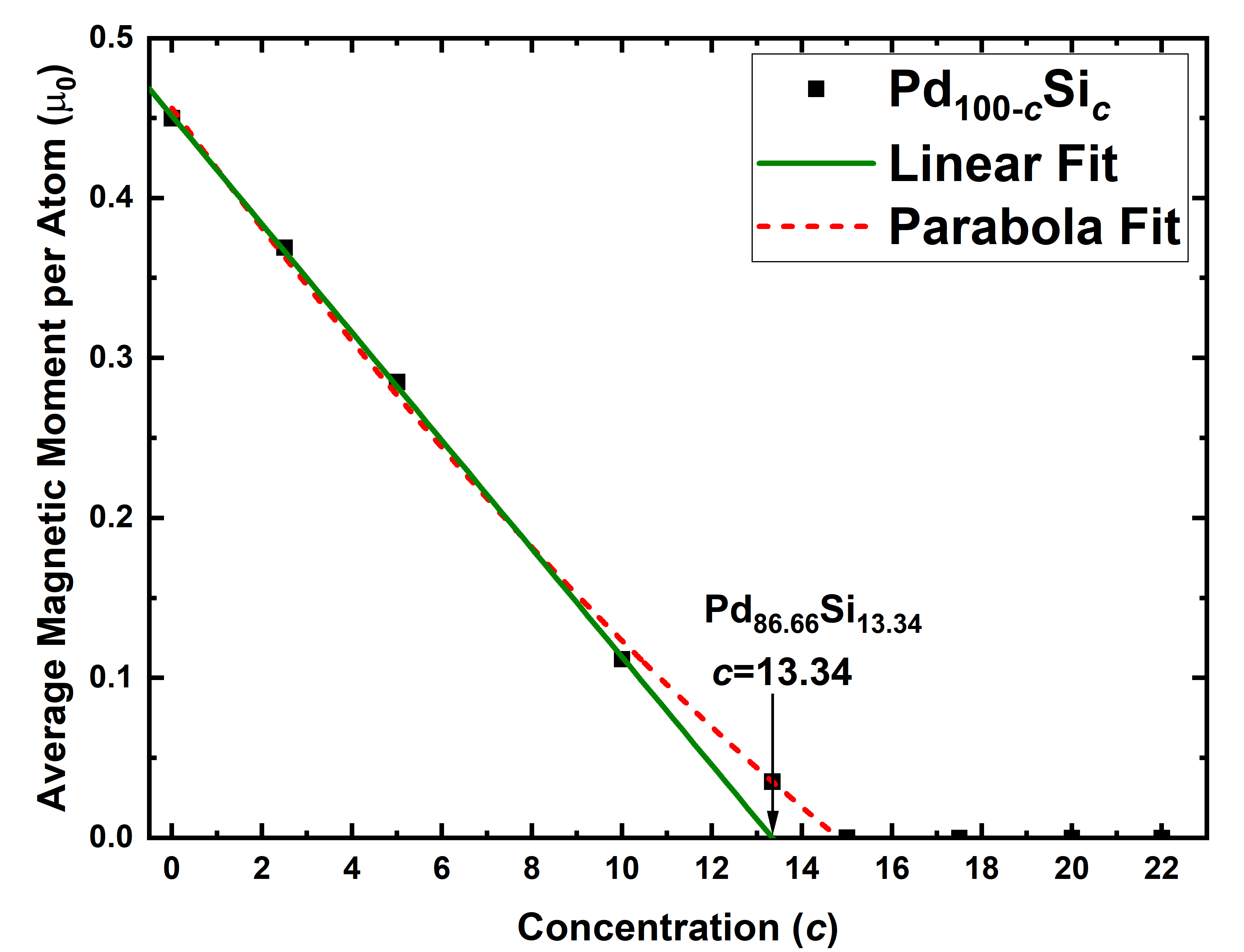}
\caption{Average Magnetic Moment, AMM, per atom after the GO runs. The alloy $a$-Pd$_{86.66}$Si$_{13.34}$ was studied to investigate a possible linear fit to our magnetic results. The fit is not linear, green line, it is quadratic, red broken line. See text.}
\label{fig:Fig6}
\end{figure}

After the GO runs we calculated the Average Magnetic Moment (AMM) per atom to investigate the possible magnetism of the PdSi system. To do so spin up ($\alpha$ spins) and spin down ($\beta$ spins) densities of states were obtained for the 9 supercells and so were the areas under each curve for all the structures, and the differences in the areas were obtained; this gives the \textbf{\textit{net}} magnetic moment for the supercell. To get the normalized results we divide by the number of atoms, the same for each cell. Figures \ref{fig:Fig5} depicts all the spin up and spin down results, and it can be observed that the asymmetry diminishes (the \textbf{\textit{net}} AMM per atom tends to zero) as the concentration of silicon increases; see also Figure \ref{fig:Fig6}. This indicates that increasing the concentration of silicon balances the loose spins in pure amorphous Pd up to about 15 at \%, giving a quantitative idea of the ``defects'' present in the amorphous pure structure. The symmetry of the partial silicon contributions to the up and down densities of states leads us to conclude that there is no \textbf{\textit{net}} AMM in the silicon atoms. Moreover, the vanishing asymmetries of the $\alpha$ spins and $\beta$ spins contributions of the Pd atoms as $c$ increases, indicates that the magnetism is associated to these atoms. More detailed studies are needed to inquire into these conclusions, and to discern the origin and evolution of magnetism.

\pagebreak

We would like to speculate about the differences between the results reported herein, and those of M\"{u}ller \textit{et al}. \cite{Muller_1978} for the liquid counterpart, in the light of S\"{a}nger's explanation \cite{Sanger_1984} of the experimentally determined total magnetic susceptibility, $\chi(T,c)$. S\"{a}nger’s decomposition indicates contributions from the $d$-spin paramagnetism, $\chi_{d} (T,c)$, the $s/p$ band spin paramagnetism, $\chi_{s} (c)$, the Pd $d$-band orbital paramagnetism, $\chi_{orb} (c)$, and the total diamagnetism, $\chi_{dia} (c)$, of the constituents:

\begin{equation*}
\chi(T,c) = \chi_{d} (T,c) + \dfrac{2}{3} \chi_{s} (c) + \chi_{orb} (c) + \chi_{dia} (c)
\end{equation*}

\begin{figure}[H]
\centering
\includegraphics[width=0.85\linewidth]{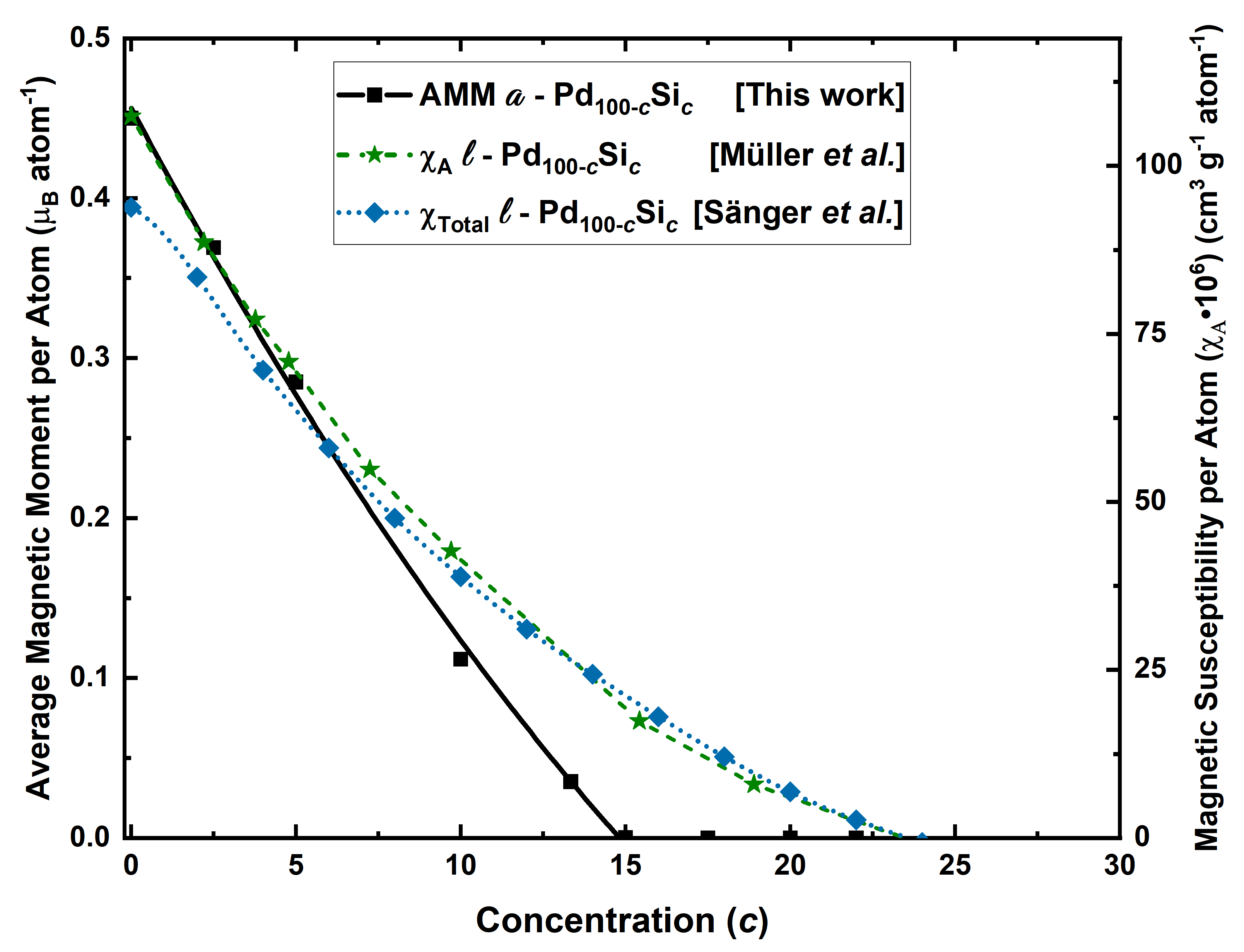}
\caption{\textit{Qualitative} comparison of our results for the AMM per atom of the solid $a$-PdSi alloys (vertical scale on the left) with the magnetic susceptibility measurements for the liquid counterpart (vertical scale on the right).}
\label{fig:Fig7}
\end{figure}

This expression indicates that the only temperature-dependent contribution to the magnetic susceptibility of the alloys appears in the $d$-contribution due to palladium, but no phase-dependent contributions are invoked.  If we were to displace the experimental susceptibility results (Figure 1 from Ref. \cite{Sanger_1984}) downwards rigidly until the green curve in Figure \ref{fig:Fig7} and the simulational results coincide at 15 at \%, one would have to conclude that the terms that S\"{a}nger invoke are relevant for low concentrations in the liquid and are not as relevant for the amorphous. In fact, based on the results of the magnetism found for the amorphous pure, solid, palladium reported elsewhere \cite{Rodriguez_2019} we believe the $d$-contribution to be more relevant and may account for higher values of magnetism in the amorphous solid PdSi alloys. However, more serious work is needed to elucidate the relevance of each contribution in the solid amorphous phases, and the why.

These na\"{i}ve conclusions should be handled with care. To our knowledge, to date no semiphenomenological calculations, like S\"{a}nger's, have been reported for solid amorphous PdSi alloys. It would be necessary to do a more rigorous estimation of any of the terms considered for the liquid alloys, that may be relevant for the solid ones, to reach reasonable conclusions.

It is important to mention that we did not study the possible magnetic cluster-like structures \cite{Handbook_2019} in the samples, which may give us an indication of the presence of nano-magnetism due to the presence of nano-domains. These structures have been mentioned in several places and should be interesting to study the feasibility of such constructions for small supercells like the ones reported here.

At first we excluded the run for $c$ = 13.34 ($a$-Pd$_{86.66}$Si$_{13.34}$) since we did not see any need to do it, but when we adjusted curves to the initial results we wanted to investigate the possible linear fit to the AMM as a function of Si concentration, to see if it was the best. When we did the run for this concentration and found that the zero intercept of the line did not occur for $c$ = 13.34 (the AMM per atom was non-zero for this value), we opted for a parabolic fit to describe our results. The subsequent concentrations were as close to zero as one can expect and in Table \ref{tab:Table1} we list the magnetic moments found for the 9 samples. The parabola fit is done for the pure and for the silicon five lowest concentrations; the linear fit is done for the pure and for the silicon three lowest concentrations.  The four highest concentrations ($c$ = 15, 17.5, 20 and 22) have an AMM essentially null, see Figure \ref{fig:Fig6} and Table \ref{tab:Table1}. Figure \ref{fig:Fig7} displays a qualitative comparison of our results and the experimental liquid results of M\"{u}ller \textit{et al}. \cite{Muller_1978}.

\begin{table}[H]
\centering
\begin{tabular}{@{}lc@{}}
\toprule
\toprule
\multicolumn{1}{c}{Alloy} & \multicolumn{1}{c}{\begin{tabular}[c]{@{}c@{}}AMM\\ $\left( \mu_{0} \right)$\end{tabular}} \\ \midrule
Pd$_{78}$Si$_{22}$        & $6.30 \times 10^{-7}$ \\
Pd$_{80}$Si$_{20}$        & $8.28 \times 10^{-7}$ \\
Pd$_{82.5}$Si$_{17.5}$    & $5.05 \times 10^{-7}$ \\
Pd$_{85}$Si$_{15}$        & $1.06 \times 10^{-4}$ \\
Pd$_{86.66}$Si$_{13.34}$  & $0.04$                \\
Pd$_{90}$Si$_{10}$        & $0.11$                \\
Pd$_{95}$Si$_{5}$         & $0.29$                \\
Pd$_{97.5}$Si$_{2.5}$     & $0.37$                \\
Pd$_{100}$                & $0.45$                \\ \bottomrule
\bottomrule
\end{tabular}
\caption{\label{tab:Table1}AMM per atom for the pure bulk, amorphous, sample of palladium and the eight alloys studied in this work. Two curve fittings are shown in Figure \ref{fig:Fig6}.}
\end{table}

Byproducts of this investigation are the results for the positions of the maxima of the first prominent peaks of the PDFs, and the coordination numbers of Si around Pd and Pd around Pd; all these are compared with some experimental results, past and recent, in what follows. An analysis of the prominent peaks of the pPDFs is presented in Table \ref{tab:Table2} where the positions of the simulated peaks for first-neighbors Pd-Si and Pd-Pd are given so the comparison with experiment can be carried out. The experimental results give 2.40 \AA\ obtained with neutron techniques (Ref. \cite{Wong_Guntherodt_book_1981}, Table 4.3, p. 71) for the first peak of the Pd-Si pPDFs and we find that the average of our 8 concentrations is 2.41, a good agreement (Table \ref{tab:Table2}). Nevertheless, an increasing subtle tendency is observed for the value of the positions of the maxima of these peaks when the concentration of Si increases. On the other hand, experimentally ``the coordination of silicon by palladium in Pd-Si glasses varies from 6 to 7 with decreasing silicon content and extrapolates to 9 for pure (hypothetical) amorphous palladium'' \cite{Wong_Guntherodt_book_1981}. Our simulations show that the coordination of Si by Pd (Z$_{\text{Si-Pd}}$) systematically increases from 8.4 for Pd$_{78}$Si$_{22}$ to 9.2 for Pd$_{97.5}$Si$_{2.5}$ when the concentration of Si decreases, which agrees with the tendency of some recent experimental results \cite{Ohkubo_2003, Nishi_1988}, as well as with the hard sphere model of Boudreaux \cite{Boudreaux_1978}, and with the Pd$_{80}$Si$_{20}$ \textit{ab-initio} value of Durandurdu \cite{Durandurdu_2012}. For the palladium surrounded by Pd (Z$_{\text{Pd-Pd}}$) we find that the coordination increases from 8.75 for Pd$_{78}$Si$_{22}$ to 11.07 for pure palladium, in contrast with the hard sphere model of Boudreaux that stays constant at around 10.5, for concentrations $10\% \leq c \leq 30\%$ \cite{Boudreaux_1978}. Compare with the extrapolated experimental value of 9 quoted in Ref. \cite{Wong_Guntherodt_book_1981}. See Table \ref{tab:Table3}.

\begin{table}[H]
\centering
\begin{tabular}{@{}lcc@{}}
\toprule
\toprule
\multicolumn{1}{c}{Alloy} & \multicolumn{1}{c}{\begin{tabular}[c]{@{}c@{}}R$_{\text{1-1}} \left(\text{\AA} \right)$\\[0.2cm] Pd-Si\end{tabular}} & \multicolumn{1}{c}{\begin{tabular}[c]{@{}c@{}}R$_{\text{1-2}} \left(\text{\AA} \right)$\\[0.2cm] Pd-Pd\end{tabular}} \\ \midrule
Pd$_{78}$Si$_{22}$        & $2.425$     & $2.765$   \\
Pd$_{80}$Si$_{20}$        & $2.425$     & $2.765$   \\
Pd$_{82.5}$Si$_{17.5}$    & $2.415$     & $2.755$   \\
Pd$_{85}$Si$_{15}$        & $2.415$     & $2.735$   \\
Pd$_{86.66}$Si$_{13.34}$  & $2.405$     & $2.745$   \\
Pd$_{90}$Si$_{10}$        & $2.405$     & $2.725$   \\
Pd$_{95}$Si$_{5}$         & $2.385$     & $2.695$   \\
Pd$_{97.5}$Si$_{2.5}$     & $2.395$     & $2.685$   \\
Pd$_{100}$                & -           & $2.685$   \\ \midrule
Average                   & $2.409$     & $2.728$   \\ \bottomrule
\bottomrule
\end{tabular}
\caption{\label{tab:Table2}Positions (R) in \AA\ for the first two prominent peaks of the pPDFs to compare with experiment. The position of the simulated first-neighbor Pd-Si peak is, on average, 2.41 \AA; the experimental value is 2.4 \AA \cite{Wong_Guntherodt_book_1981}.}
\end{table}

\begin{table}[ht]
\centering
\scalebox{0.75}{
\begin{tabular}{@{}lccccccccccccccc@{}}
\toprule
\toprule
\multicolumn{1}{c}{\multirow{2}{*}{Alloy}} & \multicolumn{3}{c}{{[}This work{]}}                                                                  & \multicolumn{3}{c}{{[}Boudreaux{]}}                                                                  & \multicolumn{3}{c}{(exp.){[}Ohkubo{]}}                                                               & \multicolumn{3}{c}{(exp.){[}Suzuki{]}}                                                               & \multicolumn{3}{c}{{[}Durandurdu{]}}                                                                 \\
\multicolumn{1}{c}{}                       & \multicolumn{1}{c}{Z$_{\text{Pd-Si}}$} & \multicolumn{1}{c}{Z$_{\text{Si-Pd}}$} & Z$_{\text{Pd-Pd}}$ & \multicolumn{1}{c}{Z$_{\text{Pd-Si}}$} & \multicolumn{1}{c}{Z$_{\text{Si-Pd}}$} & Z$_{\text{Pd-Pd}}$ & \multicolumn{1}{c}{Z$_{\text{Pd-Si}}$} & \multicolumn{1}{c}{Z$_{\text{Si-Pd}}$} & Z$_{\text{Pd-Pd}}$ & \multicolumn{1}{c}{Z$_{\text{Pd-Si}}$} & \multicolumn{1}{c}{Z$_{\text{Si-Pd}}$} & Z$_{\text{Pd-Pd}}$ & \multicolumn{1}{c}{Z$_{\text{Pd-Si}}$} & \multicolumn{1}{c}{Z$_{\text{Si-Pd}}$} & Z$_{\text{Pd-Pd}}$ \\ \midrule
Pd$_{70}$Si$_{30}$ & - & - & - & $3.29$ & $8.40$ & $10.34$ & - & - & - & - & - & - & - & - & - \\
Pd$_{78}$Si$_{22}$ & $2.40$ & $8.40$ & $8.75$ & $2.19$ & $8.26$ & $10.72$ & - & - & - & - & - & - & - & - & - \\
Pd$_{80}$Si$_{20}$ & $2.18$ & $8.19$ & $9.28$ & $2.05$ & $8.44$ & $10.22$ & - & - & - & $1.64$ & $6.56$ & $10.60$ & $2.17$ & $8.70$ & $10.77$ \\
Pd$_{82}$Si$_{18}$ & - & - & - & $1.78$ & $8.36$ & $10.49$ & $1.80$ & $8.00$ & $10.60$ & - & - & - & - & - & - \\
Pd$_{82.5}$Si$_{17.5}$ & $1.88$ & $8.82$ & $9.53$ & - & - & - & - & - & - & $1.38$ & $7.58$ & $10.60$ & - & - & - \\
Pd$_{85}$Si$_{15}$ & $1.54$ & $8.85$ & $9.70$ & $1.44$ & $8.46$ & $10.59$ & - & - & - & - & - & - & - & - & - \\
Pd$_{86.66}$Si$_{13.34}$ & $1.34$ & $8.88$ & $9.90$ & - & - & - & - & - & - & - & - & - & - & - & - \\
Pd$_{90}$Si$_{10}$ & $1.01$ & $8.91$ & $10.29$ & $0.85$ & $8.21$ & $10.65$ & - & - & - & - & - & - & - & - & - \\
Pd$_{95}$Si$_{5}$ & $0.49$ & $9.00$ & $10.48$ & - & - & - & - & - & - & - & - & - & - & - & - \\
Pd$_{97.5}$Si$_{2.5}$ & $0.26$ & $9.17$ & $10.76$ & - & - & - & - & - & - & - & - & - & - & - & - \\
Pd$_{100}$ & - & - & $11.07$ & - & - & - & - & - & - & - & - & - & - & - & - \\ \bottomrule
\bottomrule
\end{tabular}
}
\caption{\label{tab:Table3}Some coordination numbers (Z) for PdSi alloys, experimental \cite{Wong_Guntherodt_book_1981, Ohkubo_2003, Nishi_1988} and simulational \cite{Nishi_1988, Boudreaux_1978}, compared to our results. }
\end{table}

\pagebreak

\section*{Conclusions}

After having found magnetic properties in amorphous bulk palladium \cite{Rodriguez_2019}, we then decided to consider possible manifestations of magnetism in systems based on Pd that would give certainty to our findings in the amorphous and in the amorphous/nano-porous phases. Hence we studied the contamination of amorphous Pd with the isotopes H, D and T, $a$-Pd (H/D/T)$_{x}$ (where $x$ is the ratio of the contaminants), and found that, for ratios $x < 1$, $a$-Pd(H/D/T)$_{x}$ is magnetic \cite{Rodriguez_supercond_2021}. Then the next step was to study amorphous Pd$_{100-c}$Si$_{c}$ for $c$ less than or equal to 22; the results are reported herein.

It is clear that the amorphicity in Pd is responsible for the magnetism in all these materials and a more systematic study, both experimental and computational, should reveal new materials, based on palladium, that are magnetic. Also, a more detailed study is needed to clarify the nature of these magnetic properties and to enquire into the possible existence of spin-glass domains, or spin-glass clusters at the nano level.  This is underway.

Since the liquid Pd-Si alloys display magnetism, and since the structural characteristics of liquids and amorphous metallic alloys are somewhat similar, it was expected that the solid, glassy metals, $a$-Pd$_{100-c}$Si$_{c}$ should display magnetism, and they do, as shown in this paper. Other studies of Pd-based alloys are in order to investigate how wide-spread these phenomena are and to identify the type of magnetic ordering that occurs in these alloys (See reference \cite{Handbook_2019}, chapter 20 for an analysis of the variety of magnetic phenomena; in particular, magnetism in glass clusters).

A collateral inference of our studies is the otherwise evident conclusion that \textit{ab initio} studies better describe the topological aspects (position of the nearest peaks) of the structure and better describe the quantum mechanical nature of the chemical bonding.

\bibliography{PdSi}

\begin{thebibliography}{10}
\urlstyle{rm}
\expandafter\ifx\csname url\endcsname\relax
  \def\url#1{\texttt{#1}}\fi
\expandafter\ifx\csname urlprefix\endcsname\relax\def\urlprefix{URL }\fi
\expandafter\ifx\csname doiprefix\endcsname\relax\def\doiprefix{DOI: }\fi
\providecommand{\bibinfo}[2]{#2}
\providecommand{\eprint}[2][]{\url{#2}}

\bibitem{Klement_1960}
\bibinfo{author}{Klement, W.}, \bibinfo{author}{Willens, R.~H.} \&
  \bibinfo{author}{P., D.}
\newblock \bibinfo{journal}{\bibinfo{title}{Non-crystalline {S}tructure in
  {S}olidified {G}old{\textendash}{S}ilicon {A}lloys}}.
\newblock {\emph{\JournalTitle{Nature}}} \textbf{\bibinfo{volume}{187}},
  \bibinfo{pages}{869--870}, \doiprefix\url{10.1038/187869b0}
  (\bibinfo{year}{1960}).

\bibitem{Duwez_1965}
\bibinfo{author}{Duwez, P.}, \bibinfo{author}{Willens, R.~H.} \&
  \bibinfo{author}{Crewdson, R.~C.}
\newblock \bibinfo{journal}{\bibinfo{title}{Amorphous phase in
  palladium{\textemdash}silicon alloys}}.
\newblock {\emph{\JournalTitle{Journal of Applied Physics}}}
  \textbf{\bibinfo{volume}{36}}, \bibinfo{pages}{2267--2269},
  \doiprefix\url{10.1063/1.1714461} (\bibinfo{year}{1965}).

\bibitem{Duwez_Guntherodt_book_1981}
\bibinfo{author}{Duwez, P.}
\newblock \emph{\bibinfo{title}{Glassy {M}etals {I}}}, chap.
  \bibinfo{chapter}{2. Metallic Glasses-Historical Background},
  \bibinfo{pages}{19--23}.
\newblock Topics in Applied Physics (\bibinfo{publisher}{Springer-Verlag,
  Berlin Heidelberg}, \bibinfo{year}{1981}).

\bibitem{Suryanarayana_Book_2018}
\bibinfo{author}{Suryanarayana, C.} \& \bibinfo{author}{Inoue, A.}
\newblock \emph{\bibinfo{title}{Bulk {M}etallic {G}lasses}}
  (\bibinfo{publisher}{Taylor and Francis Group, CRC Press},
  \bibinfo{year}{2018}), \bibinfo{edition}{second} edn.

\bibitem{Amorphology}
\bibinfo{title}{\textit{{A}morphology {I}nc.}, {S}pin-off of {NASA} that
  produces devices based on {BMG.} {https://www.amorphology.com}}
  (\bibinfo{year}{2021}).

\bibitem{Hofmann_2016}
\bibinfo{author}{Hofmann, D.~C.} \emph{et~al.}
\newblock \bibinfo{journal}{\bibinfo{title}{Castable bulk metallic glass strain
  wave gears: Towards decreasing the cost of high-performance robotics}}.
\newblock {\emph{\JournalTitle{Scientific Reports}}}
  \textbf{\bibinfo{volume}{6}}, \bibinfo{pages}{37773},
  \doiprefix\url{10.1038/srep37773} (\bibinfo{year}{2016}).

\bibitem{Nishiyama_2012}
\bibinfo{author}{Nishiyama, N.} \emph{et~al.}
\newblock \bibinfo{journal}{\bibinfo{title}{The world{\textquotesingle}s
  biggest glassy alloy ever made}}.
\newblock {\emph{\JournalTitle{Intermetallics}}} \textbf{\bibinfo{volume}{30}},
  \bibinfo{pages}{19--24}, \doiprefix\url{10.1016/j.intermet.2012.03.020}
  (\bibinfo{year}{2012}).

\bibitem{Handbook_2019}
\bibinfo{editor}{Musgraves, J.~D.}, \bibinfo{editor}{Hu, J.} \&
  \bibinfo{editor}{Calvez, L.} (eds.) \emph{\bibinfo{title}{Springer {H}andbook
  of {G}lass}} (\bibinfo{publisher}{Springer International Publishing},
  \bibinfo{year}{2019}).

\bibitem{Cohen_1961}
\bibinfo{author}{Cohen, M.~H.} \& \bibinfo{author}{Turnbull, D.}
\newblock \bibinfo{journal}{\bibinfo{title}{Composition requirements for glass
  formation in metallic and ionic systems}}.
\newblock {\emph{\JournalTitle{Nature}}} \textbf{\bibinfo{volume}{189}},
  \bibinfo{pages}{131--132}, \doiprefix\url{10.1038/189131b0}
  (\bibinfo{year}{1961}).

\bibitem{Turnbull_1969}
\bibinfo{author}{Turnbull, D.}
\newblock \bibinfo{journal}{\bibinfo{title}{Under what conditions can a glass
  be formed?}}
\newblock {\emph{\JournalTitle{Contemporary Physics}}}
  \textbf{\bibinfo{volume}{10}}, \bibinfo{pages}{473--488},
  \doiprefix\url{10.1080/00107516908204405} (\bibinfo{year}{1969}).

\bibitem{Rodriguez_2019}
\bibinfo{author}{Rodr{\'{\i}}guez, I.}, \bibinfo{author}{Valladares, R.~M.},
  \bibinfo{author}{Hinojosa-Romero, D.}, \bibinfo{author}{Valladares, A.} \&
  \bibinfo{author}{Valladares, A.~A.}
\newblock \bibinfo{journal}{\bibinfo{title}{Emergence of magnetism in bulk
  amorphous palladium}}.
\newblock {\emph{\JournalTitle{Physical Review B}}}
  \textbf{\bibinfo{volume}{100}}, \bibinfo{pages}{024422},
  \doiprefix\url{10.1103/PhysRevB.100.024422} (\bibinfo{year}{2019}).

\bibitem{Rodriguez_supercond_2021}
\bibinfo{author}{Rodr{\'{\i}}guez, I.}, \bibinfo{author}{Valladares, R.~M.},
  \bibinfo{author}{Valladares, A.}, \bibinfo{author}{Hinojosa-Romero, D.} \&
  \bibinfo{author}{Valladares, A.~A.}
\newblock \bibinfo{title}{Superconductivity versus {M}agnetism in the
  {A}morphous {P}alladium “ides” ({P}d$_{1-x}$ ({H}/{D}/{T})$_{x}$).}
  (\bibinfo{year}{2020}).
\newblock \eprint{https://arxiv.org/abs/2012.02934}.

\bibitem{Okamoto_1993}
\bibinfo{author}{Okamoto, H.}
\newblock \bibinfo{journal}{\bibinfo{title}{{P}d-{S}i
  ({P}alladium-{S}ilicon)}}.
\newblock {\emph{\JournalTitle{Journal of Phase Equilibria}}}
  \textbf{\bibinfo{volume}{14}}, \bibinfo{pages}{536--538},
  \doiprefix\url{10.1007/bf02671983} (\bibinfo{year}{1993}).

\bibitem{Baxi_1991}
\bibinfo{author}{Baxi, H.~C.} \& \bibinfo{author}{Massalski, T.~B.}
\newblock \bibinfo{journal}{\bibinfo{title}{The {P}d{S}i (palladium-silicon)
  system}}.
\newblock {\emph{\JournalTitle{Journal of Phase Equilibria}}}
  \textbf{\bibinfo{volume}{12}}, \bibinfo{pages}{349--356},
  \doiprefix\url{10.1007/BF02649925} (\bibinfo{year}{1991}).

\bibitem{Massara_1993}
\bibinfo{author}{Massara, R.} \& \bibinfo{author}{Feschotte, P.}
\newblock \bibinfo{journal}{\bibinfo{title}{Le syst{\`{e}}me binaire
  {P}d-{S}i}}.
\newblock {\emph{\JournalTitle{Journal of Alloys and Compounds}}}
  \textbf{\bibinfo{volume}{190}}, \bibinfo{pages}{249--254},
  \doiprefix\url{10.1016/0925-8388(93)90406-D} (\bibinfo{year}{1993}).

\bibitem{Muller_1978}
\bibinfo{author}{M\"{u}ller, M.}, \bibinfo{author}{Beck, H.} \&
  \bibinfo{author}{G\"{u}ntherodt, H.~J.}
\newblock \bibinfo{journal}{\bibinfo{title}{Magnetic {P}roperties of {L}iquid
  {P}d, {S}i, and {P}d-{S}i {A}lloys}}.
\newblock {\emph{\JournalTitle{Physical Review Letters}}}
  \textbf{\bibinfo{volume}{41}}, \bibinfo{pages}{983--987},
  \doiprefix\url{10.1103/PhysRevLett.41.983} (\bibinfo{year}{1978}).

\bibitem{Sanger_1984}
\bibinfo{author}{S\"{a}nger, W.}
\newblock \bibinfo{journal}{\bibinfo{title}{On the magnetic susceptibility of
  the liquid {P}d$_{1-x}$ {S}i$_{x}$ alloy system at 1825 {K}}}.
\newblock {\emph{\JournalTitle{Zeitschrift f\"{u}r Physik B Condensed Matter}}}
  \textbf{\bibinfo{volume}{55}}, \bibinfo{pages}{13--16},
  \doiprefix\url{10.1007/BF01307494} (\bibinfo{year}{1984}).

\bibitem{Fukunaga_1981}
\bibinfo{author}{Fukunaga, T.} \& \bibinfo{author}{Suzuki, K.}
\newblock \bibinfo{journal}{\bibinfo{title}{Radial distribution functions of
  {P}d-{S}i alloy glasses by pulsed neutron total scattering measurements and
  geometrical structure relaxation simulations.}}
\newblock {\emph{\JournalTitle{Science Reports of the Research Institutes}}}
  \textbf{\bibinfo{volume}{29}}, \bibinfo{pages}{153--175}
  (\bibinfo{year}{1981}).

\bibitem{Wong_Guntherodt_book_1981}
\bibinfo{author}{Wong, J.}
\newblock \emph{\bibinfo{title}{{G}lassy {M}etals {I}}}, chap.
  \bibinfo{chapter}{4. EXAFS Studies of Metallic Glasses},
  \bibinfo{pages}{45--77}.
\newblock Topics in Applied Physics (\bibinfo{publisher}{Springer-Verlag,
  Berlin Heidelberg}, \bibinfo{year}{1981}).

\bibitem{Valladares_2008}
\bibinfo{author}{Valladares, A.~A.}
\newblock \emph{\bibinfo{title}{{G}lass {M}aterials {R}esearch {P}rogress}},
  chap. \bibinfo{chapter}{6. A New Approach to the ab initio Generation of
  Amorphous Semiconducting Structures. Electronic and Vibrational Studies},
  \bibinfo{pages}{61--123} (\bibinfo{publisher}{Nova Science Publishers, New
  York}, \bibinfo{year}{2008}).

\bibitem{Louzguine_Luzgin_2012}
\bibinfo{author}{Louzguine-Luzgin, D.~V.} \emph{et~al.}
\newblock \bibinfo{journal}{\bibinfo{title}{Atomic {S}tructure {C}hanges and
  {P}hase {T}ransformation {B}ehavior in {P}d{\textendash}{S}i {B}ulk
  {G}lass-{F}orming {A}lloy}}.
\newblock {\emph{\JournalTitle{Intermetallics}}} \textbf{\bibinfo{volume}{20}},
  \bibinfo{pages}{135--140}, \doiprefix\url{10.1016/j.intermet.2011.08.022}
  (\bibinfo{year}{2012}).

\bibitem{Clark_2005}
\bibinfo{author}{Clark, S.~J.} \emph{et~al.}
\newblock \bibinfo{journal}{\bibinfo{title}{First principles methods using
  {CASTEP}}}.
\newblock {\emph{\JournalTitle{Zeitschrift f\"{u}r Kristallographie -
  Crystalline Materials}}} \textbf{\bibinfo{volume}{220}},
  \bibinfo{pages}{567--570}, \doiprefix\url{10.1524/zkri.220.5.567.65075}
  (\bibinfo{year}{2005}).

\bibitem{Biovia}
\bibinfo{title}{Dassault {S}yst{\`{e}}mes {BIOVIA}. {M}aterials {S}tudio 2016:
  {F}orcite, {R}eflex, {CASTEP}.} (\bibinfo{year}{2016}).

\bibitem{Perdew_1996}
\bibinfo{author}{Perdew, J.~P.}, \bibinfo{author}{Burke, K.} \&
  \bibinfo{author}{Ernzerhof, M.}
\newblock \bibinfo{journal}{\bibinfo{title}{{G}eneralized {G}radient
  {A}pproximation {M}ade {S}imple}}.
\newblock {\emph{\JournalTitle{Physical Review Letters}}}
  \textbf{\bibinfo{volume}{77}}, \bibinfo{pages}{3865--3868},
  \doiprefix\url{10.1103/PhysRevLett.77.3865} (\bibinfo{year}{1996}).

\bibitem{Ohkubo_2003}
\bibinfo{author}{Ohkubo, T.} \& \bibinfo{author}{Hirotsu, Y.}
\newblock \bibinfo{journal}{\bibinfo{title}{Electron diffraction and
  high-resolution electron microscopy study of an amorphous
  {P}d$_{82}${S}i$_{18}$ alloy with nanoscale phase separation}}.
\newblock {\emph{\JournalTitle{Physical Review B}}}
  \textbf{\bibinfo{volume}{67}}, \bibinfo{pages}{094201},
  \doiprefix\url{10.1103/PhysRevB.67.094201} (\bibinfo{year}{2003}).

\bibitem{Correlation_2021}
\bibinfo{author}{Rodr{\'{\i}}guez, I.} \emph{et~al.}
\newblock \bibinfo{journal}{\bibinfo{title}{Correlation: {A}n {A}nalysis {T}ool
  for {L}iquids and for {A}morphous {S}olids}}.
\newblock {\emph{\JournalTitle{Journal of Open Source Software}}}
  \textbf{\bibinfo{volume}{6}}, \bibinfo{pages}{2976},
  \doiprefix\url{10.21105/joss.02976} (\bibinfo{year}{2021}).

\bibitem{Alvarez_2003}
\bibinfo{author}{Alvarez, F.} \& \bibinfo{author}{Valladares, A.~A.}
\newblock \bibinfo{journal}{\bibinfo{title}{First-principles {S}imulations of
  {A}tomic {N}etworks and {O}ptical {P}roperties of {A}morphous {S}i{N}$_{x}$
  alloys}}.
\newblock {\emph{\JournalTitle{Physical Review B}}}
  \textbf{\bibinfo{volume}{68}}, \bibinfo{pages}{205203},
  \doiprefix\url{10.1103/PhysRevB.68.205203} (\bibinfo{year}{2003}).

\bibitem{Waseda_1980}
\bibinfo{author}{Waseda, Y.}
\newblock \emph{\bibinfo{title}{The {S}tructure of {N}on-{C}rystalline
  {M}aterials: {L}iquids and {A}morphous {S}olids.}}
  (\bibinfo{publisher}{McGraw-Hill International Book Co},
  \bibinfo{year}{1980}).

\bibitem{Waseda_1975}
\bibinfo{author}{Waseda, Y.} \& \bibinfo{author}{Masumoto, T.}
\newblock \bibinfo{journal}{\bibinfo{title}{Structural study of an amorphous
  {P}d$_{80}${\textendash}{S}i$_{20}$ alloy by {X}-{R}ay fourier analysis}}.
\newblock {\emph{\JournalTitle{Physica Status Solidi (a)}}}
  \textbf{\bibinfo{volume}{31}}, \bibinfo{pages}{477--482},
  \doiprefix\url{10.1002/pssa.2210310217} (\bibinfo{year}{1975}).

\bibitem{The_Little_Prince}
\bibinfo{author}{de~Saint-Exup\'{e}ry, A.}
\newblock \emph{\bibinfo{title}{Le {P}etit {P}rince [{T}he {L}ittle {P}rince]}}
  (\bibinfo{publisher}{Harvest Book Div. Harcourt Inc.}, \bibinfo{year}{1943}).

\bibitem{Nishi_1988}
\bibinfo{author}{Nishi, Y.}, \bibinfo{author}{Harano, H.},
  \bibinfo{author}{Fukunaga, T.} \& \bibinfo{author}{Suzuki, K.}
\newblock \bibinfo{journal}{\bibinfo{title}{Effect of peening on structure and
  volume in a liquid-quenched {P}d$_{0.835}${S}i$_{0.165}$ glass}}.
\newblock {\emph{\JournalTitle{Physical Review B}}}
  \textbf{\bibinfo{volume}{37}}, \bibinfo{pages}{2855--2860},
  \doiprefix\url{10.1103/PhysRevB.37.2855} (\bibinfo{year}{1988}).

\bibitem{Boudreaux_1978}
\bibinfo{author}{Boudreaux, D.~S.}
\newblock \bibinfo{journal}{\bibinfo{title}{Theoretical studies on structural
  models of metallic glass alloys}}.
\newblock {\emph{\JournalTitle{Physical Review B}}}
  \textbf{\bibinfo{volume}{18}}, \bibinfo{pages}{4039--4047},
  \doiprefix\url{10.1103/PhysRevB.18.4039} (\bibinfo{year}{1978}).

\bibitem{Durandurdu_2012}
\bibinfo{author}{Durandurdu, M.}
\newblock \bibinfo{journal}{\bibinfo{title}{Ab initio modeling of metallic
  {P}d$_{80}${S}i$_{20}$ glass}}.
\newblock {\emph{\JournalTitle{Computational Materials Science}}}
  \textbf{\bibinfo{volume}{65}}, \bibinfo{pages}{44--47},
  \doiprefix\url{10.1016/j.commatsci.2012.06.040} (\bibinfo{year}{2012}).

\end{thebibliography}

\section*{Figure Legends}

\noindent \textbf{Figure 1.} Region of the Pd-Si phase diagram near the eutectic point. The atomic concentrations deployed are $0 \leq c \leq 40$ at. \%. This figure is a mathematical adjustment to the experimental values contained in References \cite{Okamoto_1993, Baxi_1991, Massara_1993}.

\noindent \textbf{Figure 2.} Total Pair Distribution Functions, tPDFs, for the 8 alloys studied in this work and for the pure bulk palladium sample. The bimodal character of the second peak, typical of amorphous Pd, gradually disappears as the silicon concentration increases. The tPDFs were calculated using \texttt{Correlation}, an open-source software developed by Rodr\'{i}guez \textit{et al.} \cite{Correlation_2021}.

\noindent \textbf{Figure 3.} Comparison of the Pd-Pd pPDF for the Pd$_{80}$Si$_{20}$ sample with the partial experimental result of Masumoto (in Waseda's book \cite{Waseda_1980}) \cite{Waseda_1975}. In the inset notice the bimodal nature of the second peak, reminiscent of the bimodal shape of the snake profile that swallowed an elephant in Le Petit Prince \cite{The_Little_Prince}.

\noindent \textbf{Figure 4.} Comparison of the XRD obtained from the experimental results of Ref. \cite{Duwez_1965} ($c = 15$ at. \%), green line, with our simulations, dark solid profile. See text.

\noindent \textbf{Figure 5.} Spin up ($\alpha$ spins) and spin down ($\beta$ spins) densities of states for the 9 supercells. (a) Total densities of states (b) Palladium partial densities of states (c) Silicon partial densities of states. The \textbf{\textit{net}} magnetic moment tends to zero as the silicon concentration increases.

\noindent \textbf{Figure 6.} Average Magnetic Moment, AMM, per atom after the GO runs. The alloy $a$-Pd$_{86.66}$Si$_{13.34}$ was studied to investigate a possible linear fit to our magnetic results. The fit is not linear, green line, it is quadratic, red broken line. See text.

\noindent \textbf{Figure 7.} \textit{Qualitative} comparison of our results for the AMM per atom of the solid $a$-PdSi alloys (vertical scale on the left) with the magnetic susceptibility measurements for the liquid counterpart (vertical scale on the right).

\pagebreak

\section*{Acknowledgements}

I.R. thanks PAPIIT, DGAPA-UNAM for his postdoctoral fellowship. D.H.-R. acknowledges Consejo Nacional de Ciencia y Tecnología (CONACyT) for supporting his graduate studies. A.A.V., R.M.V. and A.V. thank DGAPA-UNAM (PAPIIT) for continued financial support to carry out research project under Grant No. IN116520. M.T. Vázquez and O. Jiménez provided the information requested. A. Lopez and A. Pompa assisted with the technical support and maintenance of the computing unit at IIM-UNAM. Simulations were partially carried at the Computing Center of DGTIC-UNAM.

\section*{Author contributions statement}

A.A.V., A.V. and R.M.V. conceived this research and designed it with the participation of I.R. and D.H.-R. I.R. did all the simulations. All authors discussed and analyzed the results. A.A.V. wrote the first draft and the other authors enriched the manuscript.

\section*{Additional information}

\textbf{Competing interests:} The authors declare no competing interests.

\end{document}